\documentclass[11pt]{article}
\usepackage{amssymb,amsmath,amsfonts}
\usepackage{graphicx}
\usepackage{graphics}
\usepackage{epsfig}


\begin{document}

\title{TOPOLOGICAL DEFECTS AND SCALAR FIELD MODES\\
IN COSMOLOGICAL BACKGROUNDS}
\author{A. A. Saharian$^1$\thanks{%
Corresponding author, E-mail: saharian@ysu.am }, G. V. Mirzoyan$^{1,2}$, G.
H. Harutyunyan$^1$, and R. M. Avagyan$^{1,3}$ \vspace{0.3cm} \\
\textit{$^1$Institute of Physics, Yerevan State University, }\\
\textit{1 Alex Manoogian Street, 0025 Yerevan, Armenia } \vspace{0.3cm}\\
\textit{$^2$ CANDLE Synchrotron Research Institute, } \\
\textit{31 Acharyan Street, 0040 Yerevan, Armenia}\vspace{0.3cm}\\
\textit{$^3$ Institute of Applied Problems of Physics NAS RA, } \\
\textit{25 Hr. Nersisyan Street, 0014 Yerevan, Armenia}}
\maketitle

\begin{abstract}
We study topological defects with a general structure in higher-dimensional
cosmological backgrounds described by a set of angle deficit parameters. As
special cases, they include higher-dimensional generalizations of cosmic
strings and global monopoles. The corresponding complete set of mode
functions is presented for a massive scalar field with a general curvature
coupling parameter. For general scale factors and radial functions in the
line element, the angular parts of the scalar modes are expressed in terms
of associated Legendre functions. De Sitter and Milne universes are
considered as examples of cosmological expansion. For the de Sitter bulk, we
present the time-dependent parts of the mode functions in inflationary,
hyperbolic, and global coordinates.
\end{abstract}

\bigskip


\bigskip

\section{Introduction}

Topological defects are expected to arise during symmetry-breaking phase
transitions in the early universe and play an important role in a wide class
of cosmological and high-energy models \cite{Vile94,Durr02}. Depending on
the nature of the symmetry group being broken, global monopoles, cosmic
strings, and domain walls can form. In particular, cosmic strings have been
and continue to be the subject of active research due to their potential
influence on the formation of large-scale cosmic structures. Objects such as
cosmic strings and global monopoles are characterized by nontrivial
spacetime topology \cite{Vile81,Barr89}, leading to angular deficit
geometries that modify both local and global properties of quantum fields
propagating in such backgrounds. The polarization of quantum vacuum for
scalar, fermionic and vector fields has been widely investigated in the
literature for both cosmic strings and global monopoles (see, e.g., the
references in \cite{Saha24Arx,Sabi22,Li26}). As local characteristics of the
vacuum state the expectation values of the field squared, energy-momentum
tensor, and the current density were considered. The effects of vacuum
polarization become particularly significant in curved and expanding
spacetimes, where the interplay between geometry, topology, and quantum
dynamics can lead to observable consequences. In particular, cosmic string
type configurations in background of de Sitter (dS) and anti-de Sitter (AdS)
spacetimes were discussed in \cite{Ghez02}-\cite{Padu16}.

The higher-dimensional physical models have attracted considerable
attention, motivated by theories beyond the Standard Model, including string
theory and braneworld scenarios. Due to the presence of extra spatial
dimensions, both Kaluza-Klein and braneworld type, the class of topological
defects in those models is more diverse. The polarization of quantum vacuum
induced by cosmic string and global monopole type topological defects in
background of dS spacetime were studied in \cite{Beze09}-\cite{Oliv24b}.
Similar studies for another maximally symmetric background geometry, AdS
spacetime, are presented in \cite{Beze12}-\cite{Oliv24c}. Another class of
higher-dimensional models corresponds to Friedmann-Robertson-Walker
(FRW)-type spacetimes generalized to $(D+1)$ dimensions. They provide a
natural framework for studying quantum fields in expanding universes with
nontrivial spatial geometry. When topological defects are present, the
spatial sections of these spacetimes acquire angular deficit structures,
which fundamentally alter the spectrum and mode structure of quantum fields.

In this work, we consider higher-dimensional cosmological backgrounds whose
spatial sections contain angular deficit topologies. We adopt a unified
metric ansatz that incorporates both global monopole-type and cosmic
string-type geometries through a set of angular deficit parameters. This
approach allows us to treat different classes of topological defects within
a single geometric framework. The canonical quantization of fields on that
backgrounds starts by the choice of a complete set of the modes for the
field. We discuss the mode functions for a scalar field with dynamics
described by the Klein-Gordon equation containing a curvature coupling term.
Exact or analytic treatments of this equation are generally possible only
for geometries with a high degree of symmetry. Nevertheless, by exploiting
the separability of the field equation in suitably chosen coordinates, it is
possible to obtain a detailed description of the mode functions and their
associated eigenvalues. Such an analysis is essential for subsequent
investigations of vacuum polarization effects, particle creation, and
expectation values of physical observables. Particular attention is paid to
the angular sector, where the presence of angular deficits leads to a
hierarchy of eigenvalue problems and modifies the effective angular momentum
spectrum. The resulting eigenfunctions are expressed in terms of associated
Legendre functions, while the separation constants enter the radial and
temporal equations as effective parameters. The temporal part is governed by
the cosmological expansion through the scale factor, whereas the radial
equation encodes the influence of spatial curvature and topological defects.

The paper is organized as follows. In Section \ref{sec:Geom}, we describe
the geometric setup and discuss the properties of the metric, Ricci tensor,
and curvature scalar. In Section \ref{sec:Modes}, complete set of modes is
presented for a scalar field with general curvature coupling parameter and
for general cases of scale factor and radial function. A special case of a
global monopole in cosmological background is discussed in Section \ref%
{sec:Glob}. Examples of cosmological expansions are considered in Section %
\ref{sec:Expand}. The main results are summarized in Section \ref{sec:Conc}.

\section{Geometric setup}

\label{sec:Geom}

We consider a $(D+1)$-dimensional cosmological spacetime with a nontrivial
spatial topology characterized by angular deficit parameters. The line
element is chosen in a form that allows for both cosmological expansion and
topological defects in the spatial sections. In comoving coordinates, the
line element is written as

\begin{equation}
ds^{2}=-dt^{2}+a^{2}(t)\left[dr^{2}+p^{2}(r)\,d\Omega_{D-1}^{2}\right].
\label{ds2}
\end{equation}
Here, $a(t)$ is the cosmological scale factor, describing the isotropic
expansion of the universe, while the function $p^{2}(r)$ determines the
spatial curvature and allows for different radial geometries. The angular
part of the line element, $d\Omega_{D-1}^{2}$, represents a generalized $%
(D-1) $-dimensional sphere that incorporates angular deficit parameters
associated with topological defects. We will take it in the form

\begin{equation}
d\Omega _{D-1}^{2}=\sum_{i,k=1}^{D-1}\gamma _{ik}d\theta _{i}d\theta
_{k}=\alpha _{1}^{2}d\theta _{1}^{2}+\sum_{i=2}^{D-1}\alpha _{i}^{2}\left(
\prod_{l=1}^{i-1}\sin ^{2}\theta _{l}\right) d\theta _{i}^{2}.  \tag{ }
\label{angul}
\end{equation}%
The constants $\alpha _{i}^{2}$ encode possible angular deficits and reduce
to unity in the absence of topological defects. Different choices of these
parameters allow one to describe a variety of geometries, including global
monopole-type configurations with $\alpha _{1}=\alpha _{2}=\cdots =\alpha
_{D-1}$, and cosmic string type defects with $\alpha _{1}=\alpha _{2}=\cdots
=\alpha _{D-2}$ and $0<\alpha _{D-1}<1$.

In order to describe the dynamics of fields on background spacetime given by
(\ref{ds2}), we need the corresponding geometrical characteristics. For the
components of the Ricci tensor (we use the convention $R_{ik}=\partial
_{l}\Gamma _{ik}^{l}-\cdots $) one gets%
\begin{align}
R_{0}^{0}& =D\frac{\ddot{a}}{a},\;R_{1}^{1}=\frac{\ddot{a}}{a}+(D-1)\left(
\frac{\dot{a}^{2}}{a^{2}}-\frac{p^{\prime \prime }}{a^{2}p}\right) ,  \notag
\\
R_{2}^{2}& =\frac{\ddot{a}}{a}+(D-1)\frac{\dot{a}^{2}}{a^{2}}-\frac{1}{a^{2}}%
\left[ \frac{p^{\prime \prime }}{p}+(D-2)\left( \frac{p^{\prime 2}}{p^{2}}-%
\frac{\alpha _{1}^{-2}}{p^{2}}\right) \right] ,  \notag \\
R_{i}^{i}& =R_{i-1}^{i-1}+\frac{(D-i)\left( \alpha _{i-1}^{-2}-\alpha
_{i-2}^{-2}\right) }{a^{2}p^{2}\sin ^{2}\theta _{1}\cdots \sin ^{2}\theta
_{i-2}},  \label{Ricci}
\end{align}%
for $i=3,\ldots ,D$. Here, the overdot and the prime stand for the
derivatives with respect to $t$ and $r$, respectively. Note that $%
R_{D}^{D}=R_{D-1}^{D-1}$. The expression for the Ricci scalar reads
\begin{equation}
R=2D\frac{\ddot{a}}{a}+D(D-1)\frac{\dot{a}^{2}}{a^{2}}-\frac{D-1}{a^{2}}%
\left[ \frac{2p^{\prime \prime }}{p}+(D-2)\left( \frac{p^{\prime 2}}{p^{2}}-%
\frac{\alpha _{1}^{-2}}{p^{2}}\right) \right] +\frac{R_{\theta }}{a^{2}p^{2}}%
,  \label{Rscalar}
\end{equation}%
where the dependence on the angular coordinates is given by
\begin{equation}
R_{\theta }=\sum_{s=3}^{D}\sum_{i=3}^{s}\frac{(D-i)\left( \alpha
_{i-1}^{-2}-\alpha _{i-2}^{-2}\right) }{\sin ^{2}\theta _{1}\cdots \sin
^{2}\theta _{i-2}}.  \label{Rteta}
\end{equation}%
The expression for $R_{\theta }$ can also be written in the form%
\begin{equation}
R_{\theta }=\sum_{i=2}^{D-1}(D-i-1)\frac{\left( D-i\right) \left( \alpha
_{i}^{-2}-\alpha _{i-1}^{-2}\right) }{\sin ^{2}\theta _{1}\cdots \sin
^{2}\theta _{i-1}},\;D\geq 3.  \label{Rteta2}
\end{equation}%
Note that $R_{\theta }=0$ for $D=3$.

In the absence of topological defect we have $\alpha _{1}=\alpha _{2}=\cdots
=\alpha _{D-1}=1$ and the line element (\ref{ds2}) describes a general
spherically symmetric spacetime. The presence of a defect, in general,
breaks the spherical symmetry. The exception is the global monopole. Of
particular interest are cosmological models with spacetime of constant
curvature (in the absence of defects), which form the basis of standard
cosmology. For constant curvature spaces in the absence of defects the
components $R_{(0)i}^{k}$ of the Ricci tensor do not depend on the radial
coordinate. From the expression for $R_{(0)1}^{1}$ (see (\ref{Ricci})) we
conclude that
\begin{equation}
\frac{p^{\prime \prime }}{p}=-\frac{k}{\beta ^{2}},\;\beta \geq 0,\;k=0,\pm
1,  \label{pcc1}
\end{equation}%
with $\beta $ being a constant determining the curvature radius of the
space. The limit $\beta \rightarrow \infty $ corresponds to flat space with $%
p(r)=r$. For curved spaces and for $D\geq 3$, imposing the condition $%
(p^{\prime }/p)^{2}-p^{-2}=\mathrm{const}$, one has the following
possibilities:%
\begin{eqnarray}
p(r) &=&r,\;k=0,  \notag \\
p(r) &=&\beta \sin \chi ,\;k=1,  \notag \\
p(r) &=&\beta \sinh \chi ,\;k=-1,  \label{pcc2}
\end{eqnarray}%
with the dimensionless coordinate $\chi =r/\beta $. For $D=2$ we have an
additional solution $p(r)=\beta \cosh \chi $ for $k=1$, describing a
constant curvature space in the absence of defects. The mentioned three
possibilities for the function $p(r)$ correspond to FRW cosmological models
with the curvature parameter $k=0,\pm 1$. Note that for these spaces%
\begin{equation}
\frac{p^{\prime 2}}{p^{2}}-\frac{1}{p^{2}}=-\frac{k}{\beta ^{2}}.
\label{pcc3}
\end{equation}%
In the absence of defects, for the Ricci tensor corresponding to
cosmological models with maximally symmetric spaces one has (no summation
over $i$)%
\begin{equation}
R_{(0)0}^{0}=D\frac{\ddot{a}}{a},\;R_{(0)i}^{i}=\frac{\ddot{a}}{a}%
+(D-1)\left( \frac{\dot{a}^{2}}{a^{2}}+\frac{k\beta ^{-2}}{a^{2}}\right)
,\;i=1,2,\ldots ,D.  \label{Rik0}
\end{equation}%
In the presence of defects, the corresponding components have the form%
\begin{align}
R_{i}^{i}& =R_{(0)i}^{i},\;i=0,1,  \notag \\
R_{2}^{2}& =R_{1}^{1}+(D-2)\frac{\alpha _{1}^{-2}-1}{a^{2}p^{2}},  \notag \\
R_{i}^{i}& =R_{i-1}^{i-1}+\frac{(D-i)\left( \alpha _{i-1}^{-2}-\alpha
_{i-2}^{-2}\right) }{a^{2}p^{2}\sin ^{2}\theta _{1}\cdots \sin ^{2}\theta
_{i-2}}.  \label{Rikms}
\end{align}%
For the Ricci scalar we get%
\begin{equation}
R=R_{(0)}+\frac{1}{a^{2}p^{2}}\left[ \left( D-1\right) (D-2)\left( \alpha
_{1}^{-2}-1\right) +R_{\theta }\right] .  \label{Rms}
\end{equation}%
Note that for a cosmic string type defect $R_{l}^{i}=R_{(0)l}^{i}$.

\section{Scalar field modes}

\label{sec:Modes}

The dynamics of a scalar field $\varphi (x)$ is governed by the Klein-Gordon
equation with a curvature coupling:

\begin{equation}
\left( g^{ik}\nabla _{i}\nabla _{k}-m^{2}-\xi R\right) \varphi (x)=0,
\label{Feq}
\end{equation}%
where $m$ is the mass of the field, $\xi $ is the non-minimal coupling
constant. In the case of a conformally coupled field in $D$-dimensional
space one has $\xi =\xi _{D}\equiv (D-1)/(4D)$. For quantization of the
field in a given background we need a complete set of modes specified by a
set of quantum numbers $\sigma $. For the geometry described above, the mode
functions can be presented in a separated form%
\begin{equation}
\varphi _{\sigma }(x)=T(t)W(r)Y(\theta ),  \label{sep}
\end{equation}%
where $\theta =(\theta _{1},\theta _{2},\ldots ,\theta _{D-1})$. They are
normalized by the conditions
\begin{equation}
\int \ d^{D}x\sqrt{|g|}g^{00}\left[ \varphi _{\sigma }(x)\partial
_{t}\varphi _{\sigma ^{\prime }}^{\ast }(x)-\varphi _{\sigma ^{\prime
}}^{\ast }(x)\partial _{t}\varphi _{\sigma }(x)\right] =i\delta _{\sigma
\sigma ^{\prime }},  \label{norm}
\end{equation}%
where the star corresponds to complex conjugate. The symbol $\delta _{\sigma
\sigma ^{\prime }}$ is understood as Dirac delta function for continuous
components of the collective set $\sigma $ and as the Kronecker delta for
discrete components.

Plugging in the field equation for the function determining the time
dependence we get
\begin{equation}
\partial _{t}^{2}T+D\frac{\dot{a}}{a}\partial _{t}T+\left\{ m^{2}+\frac{%
\lambda ^{2}}{a^{2}}+D\xi \left[ 2\frac{\ddot{a}}{a}+(D-1)\frac{\dot{a}^{2}}{%
a^{2}}\right] \right\} T=0,  \label{Teq}
\end{equation}%
where $\lambda $ is a separation constant. From the equation (\ref{Teq}) it
follows that for any two functions $T_{1}(t)$ and $T_{2}(t)$ obeying the
equation, the relation%
\begin{equation}
T_{2}\partial _{t}T_{1}-T_{1}\partial _{t}T_{2}=\mathrm{const}\cdot a^{-D}.
\label{RelT}
\end{equation}%
holds. The normalization of the function $T(t)$ in (\ref{sep}) can be fixed
by the choice of the constant in the right-hand side of (\ref{RelT}) with $%
T_{1}=T(t)$ and $T_{2}=T^{\ast }(t)$. We will assume the normalization
condition%
\begin{equation}
T\partial _{t}T^{\ast }-T^{\ast }\partial _{t}T=ia^{-D}  \label{NormT}
\end{equation}%
Introducing a new function $v(t)$ in accordance with%
\begin{equation}
T(t)=a^{-\frac{D}{2}}v(t),  \label{vt}
\end{equation}%
the equation for it reads%
\begin{equation}
\left( \partial _{t}^{2}+\omega _{t}^{2}\right) v(t)=0,  \label{veq}
\end{equation}%
with time dependent frequency%
\begin{equation}
\omega _{t}^{2}=m^{2}+\frac{\lambda ^{2}}{a^{2}}+2D\left( \xi -\frac{1}{4}%
\right) \frac{\ddot{a}}{a}+D(D-1)\left( \xi -\xi _{D-1}\right) \frac{\dot{a}%
^{2}}{a^{2}}.  \label{omt}
\end{equation}%
In static backgrounds $\omega (t)=\mathrm{const}$ and it determines the
energy of the mode. For two solutions of (\ref{veq}) we have $v_{2}\partial
_{0}v_{1}-v_{1}\partial _{0}v_{2}=\mathrm{const}$. From (\ref{NormT}) we
obtain the normalization condition
\begin{equation}
v\partial _{t}v^{\ast }-v^{\ast }\partial _{t}v=i  \label{relv}
\end{equation}%
for the function $v(t)$.

The time dependence of the mode functions can also be formulated in terms of
conformal time coordinate $\tau $, defined by the relation $d\tau =dt/a(t)$.
Considering $T=T(\tau )$ as a function of conformal time, the corresponding
equation is written as%
\begin{equation}
\partial _{\tau }^{2}T+\left( D-1\right) \frac{a_{\tau }^{\prime }}{a}%
\partial _{\tau }T+\left\{ m^{2}a^{2}+\lambda ^{2}+D\xi \left[ 2\frac{%
a_{\tau }^{\prime \prime }}{a}+(D-3)\frac{a_{\tau }^{\prime 2}}{a^{2}}\right]
\right\} T=0,  \label{Teta}
\end{equation}%
where $a_{\tau }^{\prime }=\partial _{\tau }a$. Introducing the function $%
w=w(\tau )$ by the relation $w(\tau )=a^{(D-1)/2}T(\tau )$, the
corresponding equation takes the form%
\begin{equation}
\left( \partial _{\tau }^{2}+\omega _{\tau }^{2}\right) w(\tau )=0,
\label{weta}
\end{equation}%
with the frequency%
\begin{equation}
\omega _{\tau }^{2}=m^{2}a^{2}+\lambda ^{2}+D\left( \xi -\xi _{D}\right)
\left[ 2\frac{a_{\tau }^{\prime \prime }}{a}+(D-3)\frac{a_{\tau }^{\prime 2}%
}{a^{2}}\right] .  \label{ometa}
\end{equation}%
Note that we have the relation $w(\tau )=a^{-1/2}v(t)$. For two solutions of
the equation (\ref{weta}), $w_{1}$ and $w_{2}$, we have $w_{2}\partial
_{\tau }w_{1}-w_{1}\partial _{\tau }w_{2}=\mathrm{const}$. From (\ref{relv})
the normalization relation $w\partial _{\tau }w^{\ast }-w^{\ast }\partial
_{\tau }w=i$ is obtained for the function $w(\tau )$.

The equation for the radial function reads
\begin{eqnarray}
0 &=&W^{\prime \prime }+\left( D-1\right) \frac{p^{\prime }}{p}W^{\prime
}+\left\{ \lambda ^{2}-\frac{\gamma }{p^{2}}\right.  \notag \\
&&\left. +\left( D-1\right) \xi \left[ \frac{2p^{\prime \prime }}{p}%
+(D-2)\left( \frac{p^{\prime 2}}{p^{2}}-\frac{\alpha _{1}^{-2}}{p^{2}}%
\right) \right] \right\} W.  \label{Req}
\end{eqnarray}%
Introducing a new radial function $f(r)=p^{\frac{D-1}{2}}W$, it is written
in the form of the Schr\"{o}dinger equation
\begin{equation}
f^{\prime \prime }(r)+\left[ \lambda ^{2}-U_{\mathrm{eff}}(r)\right] f=0,
\label{feq}
\end{equation}%
with the effective potential%
\begin{equation}
U_{\mathrm{eff}}(r)=\frac{1}{p^{2}}\left[ \gamma +\frac{D-1}{4}\left(
D-3\right) \right] -\left( D-1\right) \xi \left[ \frac{2p^{\prime \prime }}{p%
}+(D-2)\left( \frac{p^{\prime 2}}{p^{2}}-\frac{1}{\alpha _{1}^{2}p^{2}}%
\right) \right] .  \label{Ueff}
\end{equation}%
In models with $p(r)=r$, the effective potential is simplified to $U_{%
\mathrm{eff}}(r)\propto 1/r^{2}$.

Finally, for the function expressing the angular dependence, the equation%
\begin{equation}
\left[ \frac{\partial _{\theta _{1}}\left( \sin ^{D-2}\theta _{1}\partial
_{\theta _{1}}\right) }{\alpha _{1}^{2}\sin ^{D-2}\theta _{1}}%
+\sum_{i=2}^{D-1}\frac{\partial _{\theta _{i}}\left( \sin ^{D-i-1}\theta
_{i}\partial _{\theta _{i}}\right) }{\alpha _{i}^{2}\sin ^{D-i-1}\theta
_{i}\prod_{l=1}^{i-1}\sin ^{2}\theta _{l}}+\gamma -\xi R_{\theta }\right] Y=0
\label{Angeq}
\end{equation}%
is obtained. The information on the time and radial dependence of the metric
enters in the angular equation through the integration constant $\gamma $.
In the absence of topological defects the space is spherically symmetric and
the angular functions are reduced to the standard hyperspherical harmonics
for a scalar field in flat spacetime.

The coordinate $\theta _{D-1}$ does not appear in the coefficients of the
equation (\ref{Angeq}) and the corresponding part in the function $Y(\theta )
$ is easily separated and a simple integration gives the dependence of the
form $Y_{(D-1)}(\theta _{D-1})=e^{in_{D-1}\theta _{D-1}}$, $n_{D-1}=0,\pm
1,\pm 2,\ldots $. The dependence on the remaining angles is presented in the
separated form%
\begin{equation}
Y(\theta )=\prod_{l=1}^{D-1}Y_{l}(\theta _{l}).  \label{Angsep}
\end{equation}%
The function $Y_{l}(\theta _{l})$, $l=1,2,\ldots ,D-2$, is a solution of the
differential equation
\begin{equation}
\left[ \frac{\partial _{\theta _{l}}\left( \sin ^{D-l-1}\theta _{l}\partial
_{\theta _{l}}\right) }{\sin ^{D-l-1}\theta _{l}}+u_{l}-\frac{c_{l}}{\sin
^{2}\theta _{l}}\right] Y_{(l)}(\theta _{l})=0,  \label{Eql}
\end{equation}%
where $u_{l}$ are the separation constants and%
\begin{equation}
c_{l}=\frac{\alpha _{l}^{2}}{\alpha _{l+1}^{2}}u_{l+1}+(D-l-2)(D-l-1)\xi
\left( \frac{\alpha _{l}^{2}}{\alpha _{l+1}^{2}}-1\right) ,  \label{cl}
\end{equation}%
with
\begin{equation}
u_{D-1}=n_{D-1}^{2},\;u_{1}=\alpha _{1}^{2}\gamma .  \label{uD1}
\end{equation}

The solution of (\ref{Eql}), regular at $\theta _{l}=0$, is expressed in
terms of the associated Legendre function of the first kind \cite{Olve10}:%
\begin{equation}
Y_{(l)}(\theta _{l})=\mathrm{const}\frac{P_{\nu _{l}}^{-\mu _{l}}(\cos
\theta _{l})}{\sin ^{\frac{D-l}{2}-1}\theta _{l}},\;l=1,\ldots ,D-2,
\label{Yl}
\end{equation}%
where%
\begin{eqnarray}
\mu _{l} &=&\sqrt{(D-l-2)^{2}/4+c_{l}},  \notag \\
\nu _{l} &=&\sqrt{(D-l-1)^{2}/4+u_{l}}-1/2.  \label{nul}
\end{eqnarray}%
For the function $P_{\nu _{l}}^{-\mu _{l}}(x_{l})$, with $x_{l}=\cos \theta
_{l}$, we have the relation
\begin{equation}
P_{\nu _{l}}^{-\mu _{l}}(-x_{l})=\cos \left[ \left( \nu _{l}-\mu _{l}\right)
\pi \right] P_{\nu _{l}}^{-\mu _{l}}(x_{l})-\frac{2}{\pi }\sin \left[ \left(
\nu _{l}-\mu _{l}\right) \pi \right] Q_{\nu _{l}}^{-\mu _{l}}(x_{l}),
\label{RelP}
\end{equation}%
where $Q_{\nu _{l}}^{-\mu _{l}}(x_{l})$ is the associated Legendre function
of the second kind. From here it follows that in order to have regularity at
$\theta _{l}=\pi $ the condition
\begin{equation}
\nu _{l}-\mu _{l}=n_{l},\;n_{l}=0,\pm 1,\pm 2,\ldots   \label{nulrel}
\end{equation}%
is required. This provides a recurrence relation between $u_{l}$ and $u_{l+1}
$. Having $u_{D-1}=n_{D-1}^{2}$, we can find all the separation constants $%
u_{D-2},\ldots ,u_{1}$, and then $\gamma =u_{1}/\alpha _{1}^{2}$. The set of
quantum numbers specifying the scalar field modes is given by $\sigma
=(\lambda ,\mathbf{n})$ with $\mathbf{n}=\left( n_{1},\ldots ,n_{D-1}\right)
$. The functions (\ref{Yl}) are rewritten in the form%
\begin{equation}
Y_{(l)}(\theta _{l})=\mathrm{const}\frac{P_{n_{l}+\mu _{l}}^{-\mu _{l}}(\cos
\theta _{l})}{\sin ^{\frac{D-l}{2}-1}\theta _{l}},\;l=1,\ldots ,D-2.
\label{Yl2}
\end{equation}%
We denote the angular modes with the set of quantum numbers $\mathbf{n}$ by $%
Y_{\mathbf{n}}(\theta )$, assuming that they are normalized by the condition%
\begin{equation}
\int d^{D-1}\theta \,\sqrt{\mathrm{det\,}\gamma _{ik}}Y_{\mathbf{n}^{\prime
}}^{\ast }(\theta )Y_{\mathbf{n}}(\theta )=\delta _{\mathbf{nn}^{\prime }}.
\label{Ynorm}
\end{equation}%
Fixing the normalizations for the functions $T(t)$ and $Y_{\mathbf{n}%
}(\theta )$, the normalization condition for the radial function is obtained
from (\ref{norm}).

The eigenvalues of $\gamma =u_{1}/\alpha _{1}^{2}$ are determined by the
angular part of the mode functions. Then, the eigenvalues of $\lambda $ are
determined from the radial equation (\ref{Req}), and the equation for the
time dependent part $T(t)$ does not contain additional quantum numbers. By
using the normalization conditions (\ref{NormT}) and (\ref{Ynorm}), from (%
\ref{norm}) we obtain the normalization condition for the radial function $%
W(r)=W_{\lambda ,\mathbf{n}}(r)$:%
\begin{equation}
\int \ dr\,p^{D-2}(r)W_{\lambda ,\mathbf{n}}(r)W_{\lambda ^{\prime },\mathbf{%
n}}^{\ast }(r)=\delta _{\lambda \lambda ^{\prime }},  \label{normW}
\end{equation}%
where on the right-hand side, integration is performed over the entire range
of allowed values {}{}of the radial coordinate.

In cosmological models with maximally symmetric spaces in the absence of
defects, the equations for the functions $T(t)$ and $Y(\theta )$ keep the
forms (\ref{Teq}) and (\ref{Angeq}), and the radial equation is simplified to%
\begin{equation}
W^{\prime \prime }+\left( D-1\right) \frac{p^{\prime }}{p}W^{\prime }+\left[
\lambda ^{2}-kD\xi \frac{D-1}{\beta ^{2}}+\frac{\left( D-1\right) (D-2)\xi
\left( 1-\alpha _{1}^{-2}\right) -\gamma }{p^{2}}\right] W=0,  \label{Reqms}
\end{equation}%
where the function $p=p(r)$ is given by (\ref{pcc2}) with $\chi =r/\beta $.

Given the complete set of modes, we can evaluate the two-point functions for
quantum scalar fields. For example, consider the Wightman function $%
G^{+}(x,x^{\prime })$, which is defined as the expectation value $%
G^{+}(x,x^{\prime })=\left\langle 0\right\vert \varphi (x)\varphi ^{\dagger
}(x^{\prime })\left\vert 0\right\rangle $ with $\left\vert 0\right\rangle $
being the vacuum state. Expanding the field operator in terms of the
complete set of modes $\{\varphi _{\sigma }(x),\varphi _{\sigma }^{\ast
}(x)\}$ and by taking into account that the vacuum state is nullified by the
action of the annihilation operator, the mode sum formula
\begin{equation}
G^{+}(x,x^{\prime })=\sum_{\sigma }\varphi _{\sigma }(x)\varphi _{\sigma
}^{\ast }(x^{\prime })  \label{Gsum}
\end{equation}%
is obtained, where $\sum_{\sigma }$ is understood as summation over discrete
quantum numbers and integration over the continuous ones. The local
characteristics of the vacuum states, bilinear in the field operator, are
obtained from the two-point functions and their derivatives in the
coincidence limit of the arguments. Of course, a renormalization procedure
is required for extraction of finite physical results. In addition to the
references cited in Introduction, vacuum polarization effects in some
special higher-dimensional backgrounds with topological defects were
considered, for example, in \cite{Beze06,Beze08,Grat17}. Quantum effects of
a cosmic string in the Schwarzschild geometry have been discussed in \cite%
{Davi88,Otte10,Otte11}.

\section{Global monopole in cosmological backgrounds}

\label{sec:Glob}

Consider a special case corresponding to a higher dimensional generalization
of a global monopole with $\alpha _{1}=\alpha _{2}=\cdots =\alpha
_{D-1}\equiv \alpha $.

\subsection{Angular modes}

The spacetime is spherically symmetric and the equation for the angular
functions is reduced to
\begin{equation}
\left( \Delta _{\theta }+\alpha ^{2}\gamma \right) Y=0,  \label{YGm}
\end{equation}%
where
\begin{equation}
\Delta _{\theta }=\frac{\partial _{\theta _{1}}\left( \sin ^{D-2}\theta
_{1}\partial _{\theta _{1}}\right) }{\sin ^{D-2}\theta _{1}}+\sum_{i=2}^{D-1}%
\frac{\partial _{\theta _{i}}\left( \sin ^{D-i-1}\theta _{i}\partial
_{\theta _{i}}\right) }{\sin ^{D-i-1}\theta _{i}\prod_{l=1}^{i-1}\sin
^{2}\theta _{l}}.  \label{Lapang}
\end{equation}%
is the angular part of the Laplace operator in $D$-dimensional flat space.
The function $Y_{\mathbf{n}}(\theta )$ is expressed in terms of the standard
hyperspherical harmonics of degree $l$, denoted here by $Y(\mathbf{n};\theta
)$, where $\mathbf{n}=(n_{1}\equiv l,n_{2},\ldots ,n_{D-1})$, $%
l=0,1,2,\ldots $, and $n_{2},n_{3},\ldots ,n_{D-1}$ are integers such that $%
-n_{D-2}\leqslant n_{D-1}\leqslant n_{D-2}$ and
\begin{equation}
0\leq n_{D-2}\leq n_{D-3}\leq \cdots \leq n_{2}\leq l.  \label{nn}
\end{equation}%
The equation for the hyperspherical harmonics reads
\begin{equation}
\Delta _{\theta }Y(\mathbf{n};\theta )=-l(l+D-2)Y(\mathbf{n};\theta ).
\label{EqY}
\end{equation}%
Combining this equation with (\ref{YGm}), we get $\gamma =l(l+D-2)/\alpha
^{2}$.

The normalization condition for the functions $Y(\mathbf{n};\theta )$ reads%
\begin{equation}
\int d\Omega \,\left\vert Y(\mathbf{n};\theta )\right\vert ^{2}=N(\mathbf{n}%
),  \label{Yint}
\end{equation}%
where $d\Omega $ is the solid angle element in Minkowski spacetime and the
expression for $N(\mathbf{n})$ can be found in \cite{Erde53}. According to
the above arrangement, the angular modes in the presence of a global
monopole are normalized by the condition (\ref{Ynorm}). By taking into
account that in the special case under consideration $\int d^{D-1}\theta \,%
\sqrt{\mathrm{det\,}\gamma _{ik}}=\alpha ^{D-1}\int d\Omega $, we get the
relation
\begin{equation}
Y_{\mathbf{n}}(\theta )=\frac{Y(\mathbf{n};\theta )}{\alpha ^{\frac{D-1}{2}}%
\sqrt{N(\mathbf{n})}}  \label{Anggm}
\end{equation}%
between the angular functions.

\subsection{Radial modes}

Here we specify the radial modes for maximally symmetric spaces in the
absence of a global monopole. The modes for zero, positive and negative
curvature spaces are considered separately.

\textbf{Flat space,} $k=0$, $p(r)=r$.

The solution of the radial equation (\ref{Reqms}) with $k=0$ and $\gamma
=l(l+D-2)/\alpha ^{2}$ reads%
\begin{equation}
W(r)=\mathrm{const}\cdot \frac{J_{a_{l}}\left( \lambda r\right) }{r^{\frac{D%
}{2}-1}},  \label{Rsolflat}
\end{equation}%
where the order of the Bessel function $J_{a_{l}}\left( \lambda r\right) $
is given by the expression
\begin{equation}
a_{l}=\frac{1}{\alpha }\sqrt{\left( l+D/2-1\right) ^{2}+\left( D-1\right)
(D-2)\left( \xi -\xi _{D-1}\right) \left( 1-\alpha ^{2}\right) }.
\label{nul2}
\end{equation}

\textbf{Positive curvature space,} $k=1$, $p(r)=\beta \sin \chi $.

The corresponding radial equation is reduced to%
\begin{equation}
\partial _{\chi }^{2}W(r)+\left( D-1\right) \cot \chi \,\partial _{\chi
}W(r)+\left[ \lambda ^{2}\beta ^{2}-D\left( D-1\right) \xi -\frac{%
a_{l}^{2}-(D-2)^{2}/4}{\sin ^{2}\chi }\right] W(r)=0.  \label{Reqpos}
\end{equation}%
The solution of this equation regular at $r=0$ is expressed as%
\begin{equation}
W(r)=\mathrm{const}\cdot \frac{P_{\nu _{+}-1/2}^{-a_{l}}(\cos \chi )}{\sin ^{%
\frac{D}{2}-1}\chi },  \label{Rsolpos}
\end{equation}%
with the notation%
\begin{equation}
\nu _{+}=\sqrt{\lambda ^{2}\beta ^{2}+D\left( D-1\right) \left( \xi _{D}-\xi
\right) }.  \label{nupl}
\end{equation}%
Combining the analog of the formula (\ref{RelP}) for the function $P_{\nu
_{+}-1/2}^{-a_{l}}(\cos \chi )$ with the regularity condition of the modes
at $\chi =\pi $, the relation $\nu _{+}-1/2-a_{l}=n$, with $n=0,\pm 1,\pm
2,\ldots $, is obtained for the order and degree of the associated Legendre
function. This relation determines the eigenvalues of the quantum number $%
\lambda =\lambda _{n}$:%
\begin{equation}
\lambda _{n}^{2}=\beta ^{-2}\left[ \left( n+a_{l}+1/2\right) ^{2}+D\left(
D-1\right) \left( \xi -\xi _{D}\right) \right] .  \label{lamn}
\end{equation}%
The radial functions are presented in the form%
\begin{equation}
W(r)=\mathrm{const}\cdot \frac{P_{n+a_{l}}^{-a_{l}}(\cos \chi )}{\sin ^{%
\frac{D}{2}-1}\chi },\;\chi =r/\beta .  \label{Rsolpos2}
\end{equation}%
Instead of $\lambda _{n}$, as a new quantum number specifying the modes, we
can take the number $n$ in the degree of the associated Legendre function
and for the set of quantum numbers we have $\sigma =(n,n_{1},\ldots
,n_{D-1}) $. The Casimir energy in the $D=3$ Einstein universe in the
presence of a cosmic string has been studied in \cite{Beze14,Beze21} for a
conformally coupled massless scalar field. This correspond to a special case
of the line element (\ref{ds2}) with $D=3$, $a(t)=\mathrm{const}$\textrm{,} $%
p(r)=\beta \sin \chi $, and $\alpha _{1}=1$.

\textbf{Negative curvature space,} $k=-1$, $p(r)=\beta \sinh \chi $, $0\leq
\chi <\infty $.

In terms of the variable $\chi $ the radial equation is written as%
\begin{equation}
\partial _{\chi }^{2}W(r)+\left( D-1\right) \coth \chi \,\partial _{\chi
}W(r)+\left[ \lambda ^{2}\beta ^{2}+D\left( D-1\right) \xi -\frac{%
a_{l}^{2}-(D-2)^{2}/4}{\sinh ^{2}\chi }\right] W(r)=0,  \label{Reqneg}
\end{equation}%
with the solution regular at $r=0$ given by%
\begin{equation}
W(r)=\mathrm{const}\cdot \frac{P_{i\nu _{-}-1/2}^{-a_{l}}(\cosh \chi )}{%
\sinh ^{\frac{D}{2}-1}\chi },  \label{Rsolneg}
\end{equation}%
where%
\begin{equation}
\nu _{-}=\sqrt{\lambda ^{2}\beta ^{2}+D\left( D-1\right) \left( \xi -\xi
_{D}\right) }.  \label{numi}
\end{equation}%
Note that for the associated Legendre function we have $P_{i\nu
_{-}-1/2}^{-a_{l}}(x)=P_{-i\nu -1/2}^{-a_{l}}(x)$ and the radial function $%
W(r)$ is real. Similar to the case of a positive curvature space, as
independent quantum numbers we can take the set $\sigma =\left( \nu
_{-},n_{1},\ldots ,n_{D-1}\right) $, with $0\leq \nu _{-}<\infty $.

\section{Examples of cosmological expansion}

\label{sec:Expand}

\subsection{de Sitter spacetime}

As an example let us discuss a spacetime which is reduced to the de Sitter
(dS) spacetime in the absence of topological defects. First we consider the
chart of dS spacetime covered by inflationary spherical coordinates. The
line element is given by (\ref{ds2}) with the scale factor $a(t)=e^{Ht}$,
where $H=\dot{a}/a$ is the Hubble constant. Introducing a new time
coordinate $\tau =-e^{-Ht}/H$, $-\infty <\tau <0$, the interval is written
in the form%
\begin{equation}
ds^{2}=\left( H\tau \right) ^{-2}\left( -d\tau ^{2}+dr^{2}+r^{2}d\Omega
_{D-1}^{2}\right) ,  \label{dsdSc}
\end{equation}%
conformally related to the geometry of a global monopole in background of
flat spacetime. For this line element we have%
\begin{equation}
\omega _{\tau }^{2}=\lambda ^{2}+\left( \frac{1}{4}-\nu _{\mathrm{S}%
}^{2}\right) \frac{1}{\tau ^{2}},\;\nu _{\mathrm{S}}\equiv \sqrt{%
D^{2}/4-D(D+1)\xi -m^{2}/H^{2}}.  \label{nuS}
\end{equation}%
The general solution of the equation (\ref{weta}) is a linear combination of
the functions $|\tau |^{1/2}H_{\nu _{\mathrm{S}}}^{(l)}(\lambda |\tau |)$
with $l=1,2$, where $H_{\nu }^{(l)}(\lambda \eta )$ are the Hankel
functions. The function $T(t)$ is presented in the form%
\begin{equation}
T(t)=|\tau |^{\frac{D}{2}}\sum_{l=1}^{2}C_{l}H_{\nu _{\mathrm{S}%
}}^{(l)}(\lambda |\tau |).  \label{TdS}
\end{equation}%
In the canonical quantization procedure for scalar fields in dS spacetime,
different choices of the ratio $C_{2}/C_{1}$ of the coefficients in (\ref%
{TdS}) correspond to different vacuum states. In particular, Bunch-Davies
vacuum state is realized by the choice $C_{2}=0$.

Next, consider the dS spacetime foliated by negative constant curvature
spatial sections (for a discussion of the relations between inflationary,
global, and hyperbolic coordinates of dS spacetime in the absence of defects
and the corresponding scalar field modes see \cite{Sasa95,Saha21}). In the
presence of a topological defect the modified line element is given by (\ref%
{ds2}) where $a(t)=\sinh (Ht)/(H\beta )$ and $p(r)=\beta \sinh (r/\beta )$.
For the conformal time we have $\tau =\beta \ln \left( \tanh \left(
Ht/2\right) \right) $, $-\infty <\tau <0$. In terms of this time coordinate
we have $a(\tau )=-1/[\beta H\sinh (\tau /\beta )]$. The time dependent
frequency in the equation (\ref{weta}) is expressed as
\begin{equation}
\omega _{\tau }^{2}=\frac{1}{\beta ^{2}}\left[ \nu _{-}^{2}-\frac{\nu _{%
\mathrm{S}}^{2}-1/4}{\sinh ^{2}(\tau /\beta )}\right] .  \label{omtau2}
\end{equation}%
The equation (\ref{Teq}) determining the time dependence of the mode
function is reduced to the equation for the associated Legendre functions
and the general solution reads%
\begin{equation}
T(t)=c_{1}P_{\nu _{\mathrm{S}}-1/2}^{i\nu _{-}}\left( \cosh (Ht)\right)
+c_{2}P_{\nu _{\mathrm{S}}-1/2}^{-i\nu _{-}}\left( \cosh (Ht)\right) .
\label{TdShyp}
\end{equation}%
The argument of the associated Legendre function is expressed in terms of
the conformal time as $\cosh \left( Ht\right) =-\coth \left( \tau /\beta
\right) $. The set of quantum numbers is given by $\sigma =(\nu _{-},\mathbf{%
n})$. The separation constant $\lambda ^{2}$ in (\ref{Teq}) is expressed in
terms of $\nu _{-}$ by the formula (\ref{numi}). The relative coefficient in
the linear combination is determined by the choice of the vacuum state. For
the geometry under consideration the conformal and adiabatic vacua coincide
and they correspond to the choice $c_{2}=0$.

The third model corresponds to dS spacetime covered by global coordinates.
In this model we have $a(t)=\cosh (Ht)/(H\beta )$ and $p(r)=\beta \sin
(r/\beta )$. The conformal time is given by $\tau =2\beta \arctan \left(
\tanh (Ht/2)\right) $ with $-\pi /2<\tau /\beta <\pi /2$. By taking into
account that $\cosh (Ht)=1/\cos \left( \tau /\beta \right) $, the line
element is written in the form%
\begin{equation}
ds^{2}=\frac{-d\tau ^{2}+dr^{2}+\beta ^{2}\sin ^{2}(r/\beta )\,d\Omega
_{D-1}^{2}}{H^{2}\beta ^{2}\cos ^{2}\left( \tau /\beta \right) }.
\label{ds2glob}
\end{equation}%
For the frequency in the equation (\ref{weta}) one gets%
\begin{equation}
\omega _{\tau }^{2}=\frac{1}{\beta ^{2}}\left[ \nu _{+}^{2}-\frac{\nu _{%
\mathrm{S}}^{2}-\frac{1}{4}}{\cos ^{2}\left( \tau /\beta \right) }\right] .
\label{omtau3}
\end{equation}%
The general solution of the equation determining the time dependent part in
the scalar field mode functions is expressed in terms of the associated
Legendre functions:
\begin{equation}
T(t)=\frac{c_{1}P_{\nu _{+}-1/2}^{-\nu _{\mathrm{S}}}(\tanh
(Ht))+c_{2}Q_{\nu _{+}-1/2}^{-\nu _{\mathrm{S}}}(\tanh (Ht))}{\cosh
^{D/2}(Ht)}.  \label{TdSsolgl}
\end{equation}%
This solution can also be written in terms of the conformal time by taking
into account that $\tanh (Ht)=\sin \left( \tau /\beta \right) $. From the
solution of the radial equation we have $\nu _{+}-1/2=n+a_{l}$, where $a_{l}$
is defined by (\ref{nul2}). In the absence of topological defect one gets $%
a_{l}=l+D/2-1$. Introducing a new quantum number $n^{\prime }=n+l$, in the
special case $D=3$ and $c_{2}=0$ we obtain the mode function widely
considered in the literature related to the quantization of a scalar field
in dS spacetime with global coordinates (see, for example, \cite{Birr82}).

\subsection{Milne universe}

In the absence of defects the Milne universe is described by the line
element (\ref{ds2}) with $\alpha _{1}=\alpha _{2}=\cdots =\alpha _{D-1}=1$, $%
a(t)=t/\beta $ and $p(r)=\beta \sinh \left( r/\beta \right) $ (for the
corresponding mode functions and vacuum polarization effects see \cite%
{Saha20,Saha22}). The spacetime is flat with negatively curved spatial
sections and the Ricci scalar is zero. At first glance, this may seem
contradictory to the fact that the curvature coupling constant $\xi $
appears in the time and radial equations. However, it is easy to see that it
disappears by redefining the separation constant $\lambda ^{2}$ in
accordance with $\lambda ^{2}\rightarrow \lambda ^{2}-D(D-1)\xi /\beta ^{2}$%
. With this redefinition, the equation for the function $T(t)$ reads
\begin{equation}
\left( \partial _{t}^{2}+\frac{D}{t}\partial _{t}+m^{2}+\frac{\lambda
^{2}\beta ^{2}}{t^{2}}\right) T(t)=0.  \label{TeqMilne}
\end{equation}%
The solution of this equation is expressed as
\begin{equation}
T(t)=\frac{\sum_{l=1}^{2}C_{l}H_{iz}^{(l)}(mt)}{t^{(D-1)/2}},\;z^{2}=\lambda
^{2}\beta ^{2}-\frac{(D-1)^{2}}{4}.  \label{TsolMilne}
\end{equation}%
Different choices of the ratio of the coefficients in (\ref{TsolMilne})
define different vacuum states. The choice $C_{1}=0$ corresponds to the
adiabatic vacuum, while for the conformal vacuum one takes $T(t)\propto
J_{-iz}(mt)/t^{(D-1)/2}$. The radial solution is given by (\ref{Rsolneg})
where, by taking into account the redefinition of $\lambda ^{2}$, one has $%
\nu _{-}=z$.

\section{Conclusion}

\label{sec:Conc}

In the present paper, we have considered cosmological backgrounds with
topological defects of a general structure. The defect is characterized by a
set of parameters $(\alpha _{1},\alpha _{2},\ldots ,\alpha _{D-1})$, which
describe the deficit in the corresponding angles. The model includes the two
most popular special cases corresponding to global monopoles and cosmic
strings. Generally, the presence of a topological defect breaks the
spherical symmetry of the background geometry. This can be seen from the
expression (\ref{Rscalar}) for the Ricci scalar depending on the angular
coordinates, as well as from the expression (\ref{Ricci}) for the components
of the Ricci tensor. The exception is the case of an isotropic global
monopole, for which the angle deficit parameters are $\alpha _{1}=\alpha
_{2}=\ldots =\alpha _{D-1}$. The expressions for the Ricci tensor and Ricci
scalar simplify further in cosmological backgrounds with maximally symmetric
spaces in the absence of topological defects.

The nontrivial topology and curvature of spacetime, induced by topological
defects, are sources of vacuum polarization in quantum field theory. The
canonical quantization procedure for fields in a given background relies on
a complete set of modes that solve the classical field equation. We studied
the modes of a scalar field in cosmological backgrounds with topological
defects. They are presented in a form with separated functions of time,
radial coordinate, and angles. The equations that determine the time and
radial dependencies of the mode functions are given by (\ref{Teq}) and (\ref%
{Req}). The angular part is presented in a form with separated factors being
functions of a single angle. These factors are expressed in terms of the
associated Legendre function of the first kind. The functions of time and
radial coordinate in the modes depend on the specific model under
consideration.

As an example of the application of the general scheme we have considered an
isotropic global monopole. In this case the model is spherically symmetric
and the angular part of the modes is expressed in terms of spherical
harmonics for a scalar field in Minkowski spacetime. Examples of radial
dependence include maximally symmetric spaces with zero, positive, and
negative curvature in the absence of the defect. The corresponding radial
functions are expressed in terms of Bessel and associated Legendre
functions. Specific examples of the cosmological expansion are discussed in
Section \ref{sec:Expand}. They include the dS spacetime described in
inflationary, hyperbolic and global coordinates, , as well as the Milne
universe. In these backgrounds, the additional constant in the linear
combination of the time-dependent factors of the mode functions is fixed by
the vacuum state choice.

\section*{Acknowledgment}

The work was supported by the grant No. 21AG-1C047 of the Higher Education
and Science Committee of the Ministry of Education, Science, Culture and
Sport RA. G. V. Mirzoyan was supported by the grant No. 21AG-1C006 of the
Higher Education and Science Committee of the Ministry of Education,
Science, Culture and Sport RA.

\end{document}